\documentstyle[sprocl,epsf]{article}

\def\be{\begin{equation}}
\def\ee{\end{equation}}
\def\bea{\begin{eqnarray}}
\def\eea{\end{eqnarray}}

\begin{document}

\title{POLARIZATION OBSERVABLES OF 
$\overrightarrow{p} \overrightarrow{p}\rightarrow pp \pi^0$ \\NEAR THRESHOLD}

\author{P. TH\"ORNGREN-ENGBLOM, H.O. MEYER, J.T. BALEWSKI, J. DOSKOW,
R.E. POLLOCK, B.VON PRZEWOSKI, T.~RINCKEL, F.~SPERISEN}

\address{Department of Physics and Indiana University 
Cyclotron Facility, Indiana University, Bloomington, IN 47405,
U.S.A.\\E-mail: meyer@iucf.indiana.edu, pia@iucf.indiana.edu}

\author{W.W. DAEHNICK, SWAPAN K. SAHA}

\address{Department of Physics, Pittsburgh University, 
Pittsburgh, PA 15260, U.S.A.}

\author{W. HAEBERLI, B. LORENTZ, F. RATHMANN, 
B. SCHWARTZ, T. WISE}

\address{Department of Physics, University of 
Wisconsin-Madison, Madison, WI 53706, U.S.A.}

\author{P.V. PANCELLA}

\address{Department of Physics, Western Michigan University, Kalamazoo,
MI 49008, U.S.A.}

\maketitle
\abstracts{With the aim to study spin dependence of pion production
near threshold, an internal target facility and a forward detector 
have been installed in the Cooler synchrotron ring at
IUCF. The detector system comprises scintillators
and wire chambers. The target consists of a thin-walled open-ended
storage cell into which polarized atomic hydrogen is injected.
Using a stored and cooled, polarized proton beam, polarization 
observables of the reaction 
$\overrightarrow{p} \overrightarrow{p}\rightarrow pp \pi^0$ 
have been studied at energies between 21 and 55 MeV above the
pion production threshold in the center of mass system. 
We report here measurements of the spin correlation parameters 
$\rm A_{xx}$ and $\rm A_{yy}$, and the analyzing power $\rm A_y$,
integrated over the pion polar angle.}

\section{Introduction}
\noindent Close to threshold for pion production in the NN system, 
the particles in the final state are produced in an Ss state,
(the two letters denote the relative angular
momentum of the two nucleons and of the pion with 
respect to the c.m., respectively). 
A measurement of $\sigma _{tot}$ of $p p \rightarrow pp \pi^0 $ at IUCF
at bombarding energies between 285 and 325 MeV~\cite{MEY92},
exceeded by a factor of five the theoretical predictions at the time,
which were based on direct production and on-shell rescattering 
mechanisms~\cite{KOLTUNREITAN}.
The energy dependence is well described by phase
space and the final state interaction between the two protons
in an effective-range approximation~\cite{MEY92}. In order to
account for the magnitude of the total cross section additional
effects from the exchange of the $\sigma$ and $\omega$ mesons
were incorporated into the calculations and thus it was considered 
that there was evidence for heavy meson exchange in the short-range 
NN interaction~\cite{LEERISKA}$^,$~\cite{HOROWITZ}.
Inclusion of the off-shell rescattering amplitude was also shown to
account for the discrepancy between experiment and 
theory~\cite{HERNOSET}.
Among the most recent theoretical work is a comprehensive model
calculation also taking into account explicitly the role of the $\Delta$ 
isobar~\cite{HANHART98}. Attempts to explain the Ss partial wave
strength in the framework of chiral perturbation have met with little 
success, so far~\cite{SATO}.
The interpretation of the data remains inconclusive whereas
the experimental results were confirmed and also extended closer to
threshold by an experiment at CELSIUS for beam energies between 
281 and 310 MeV~\cite{BONDAR}.

As the proton bombarding energy is
raised, the contributions of partial waves with a Ps or Pp final state 
become important. In order to separate these contributions and thus 
discriminate between different production mechanisms and models,
polarization observables are required. 
In this experiment, we have measured the spin-dependent 
cross section 
$\Delta \sigma _T$, the spin correlation para-meters
$A_{xx}$ and $A_{yy}$, 
and the analyzing power $A_y$ in a
study of $\overrightarrow{p} \overrightarrow{p} \rightarrow pp \pi^0$ 
at beam 
energies 325, 350, 375 and 400 MeV.
In section 2 the experiment is described. The data analysis and the
results are presented in section 3. The conclusion
and further plans are summarized in section 4.

\section{The Experiment}
\subsection{Beam and Target}
\noindent The proton beam was stack-injected into the Cooler synchrotron
ring at IUCF with filling rates of 30-100 $\mu$A/min. After filling
for 2-3 minutes acceleration to the bombarding energy of choice was done. Data
were taken during the flattop which lasted typically 5-8 minutes, after 
which the remaining beam was discarded. The beam lifetime was of the
order of 0.5 to 1 h. The cooled proton beam is characterized by its 
very good energy resolution 
$\delta p /p$ of the order of $10^{-5}$ and its exceptionally small 
emittance of 0.1 $\pi$ mm mrad~\cite{COOLER}. 

The target was a polarized atomic hydrogen gas storage cell installed 
in the A-region of the Cooler, fed by
an atomic beam source~\cite{WISE93}.  At this location of the ring the
$\beta$ functions are small ($\beta _x = 0.9$ m and $\beta _y = 1.7$ m) 
and the dispersion is practically zero. This means that the target
cell diameter can be small which helps to achieve a thicker target and
thus sufficient luminosity to measure polarization observables for total 
cross sections on the order of tens of $\mu$b.
The cell is a 12 mm diameter cylinder constructed from 25 $\mu$m thick
Teflon-coated aluminum foil. The Teflon surface preserves the
polarization during the approximately 300 wall collisions that the 
hydrogen atoms undergo. The atomic beam is injected through a second 
11 mm diameter tube with a wall thickness of 75 $\mu$m that is 
spot-welded to and intersects the storage cell at 30$^{\circ}$ to the normal. 
For background
and systematic error studies unpolarized gas can be fed into center of
the cell through a small spot-welded fitting on the 11 mm tube.
Mounted outside of the target chamber are three magnetic guide field
coils that govern the polarization direction of the target. 
The operational field strength was 0.2 - 2.0 mT, chosen such as to
minimize the perturbations of the closed orbit of the beam. 
Change of guide field direction is done every 2 s
and does not affect the magnitude of polarization~\cite{HAEBERLI}.
The time constant of the polarization reversal is $7 \pm 1$ ms~\cite{RATHMANN}
and data taking is off for 100 ms during every reversal.
Beam polarization was vertical ($\pm $y) and the target polarization 
was switched between vertical ($\pm $y) and horizontal ($\pm $x). 
Thus a total of 8 spin combinations were used.

\subsection{Forward detector}
\begin{figure}[t]
\epsfxsize=5cm
\epsfysize=5cm
\centerline{
\epsfbox{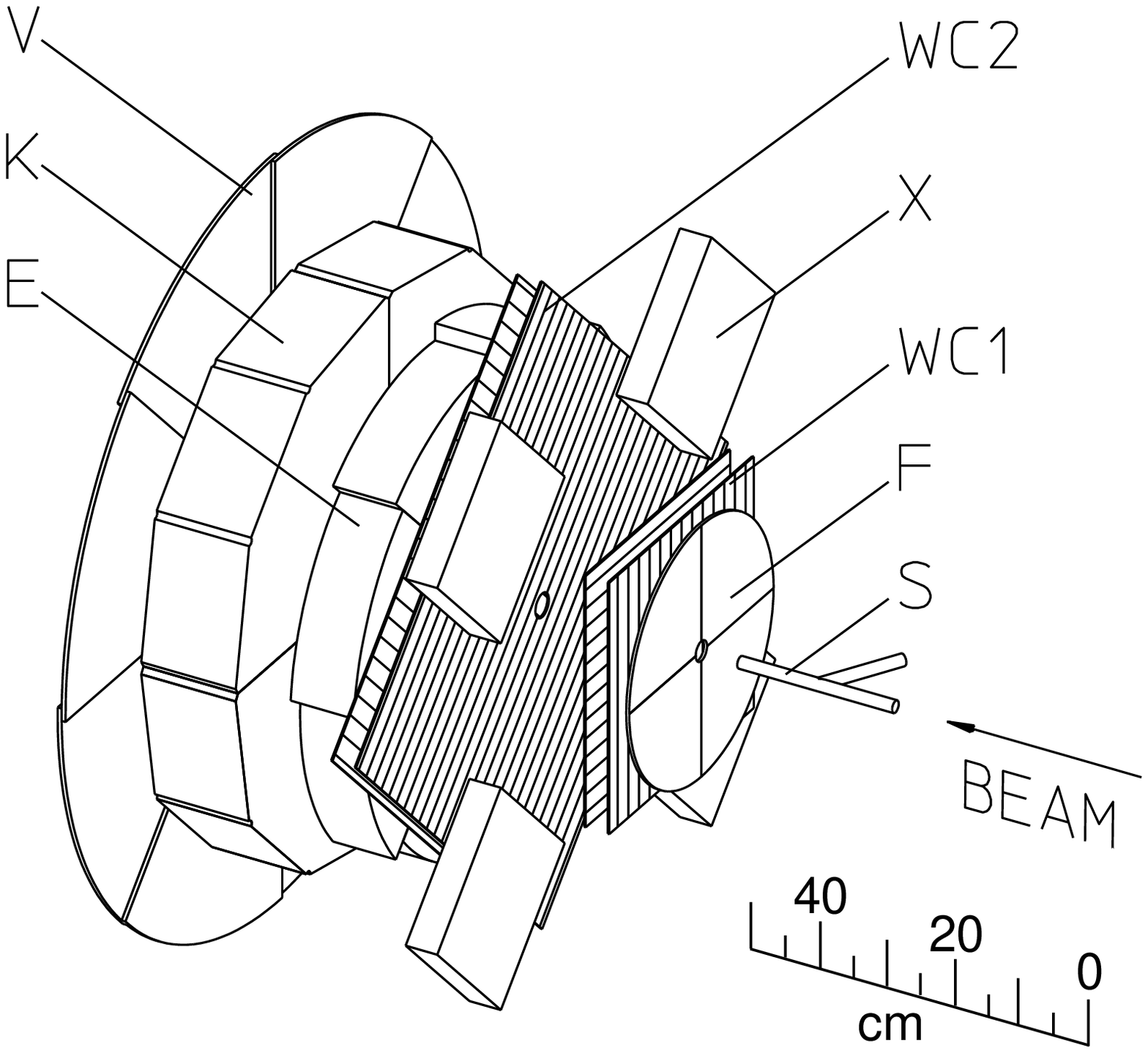}
\hspace{1cm}
\epsfxsize=5cm
\epsfysize=5cm
\epsfbox{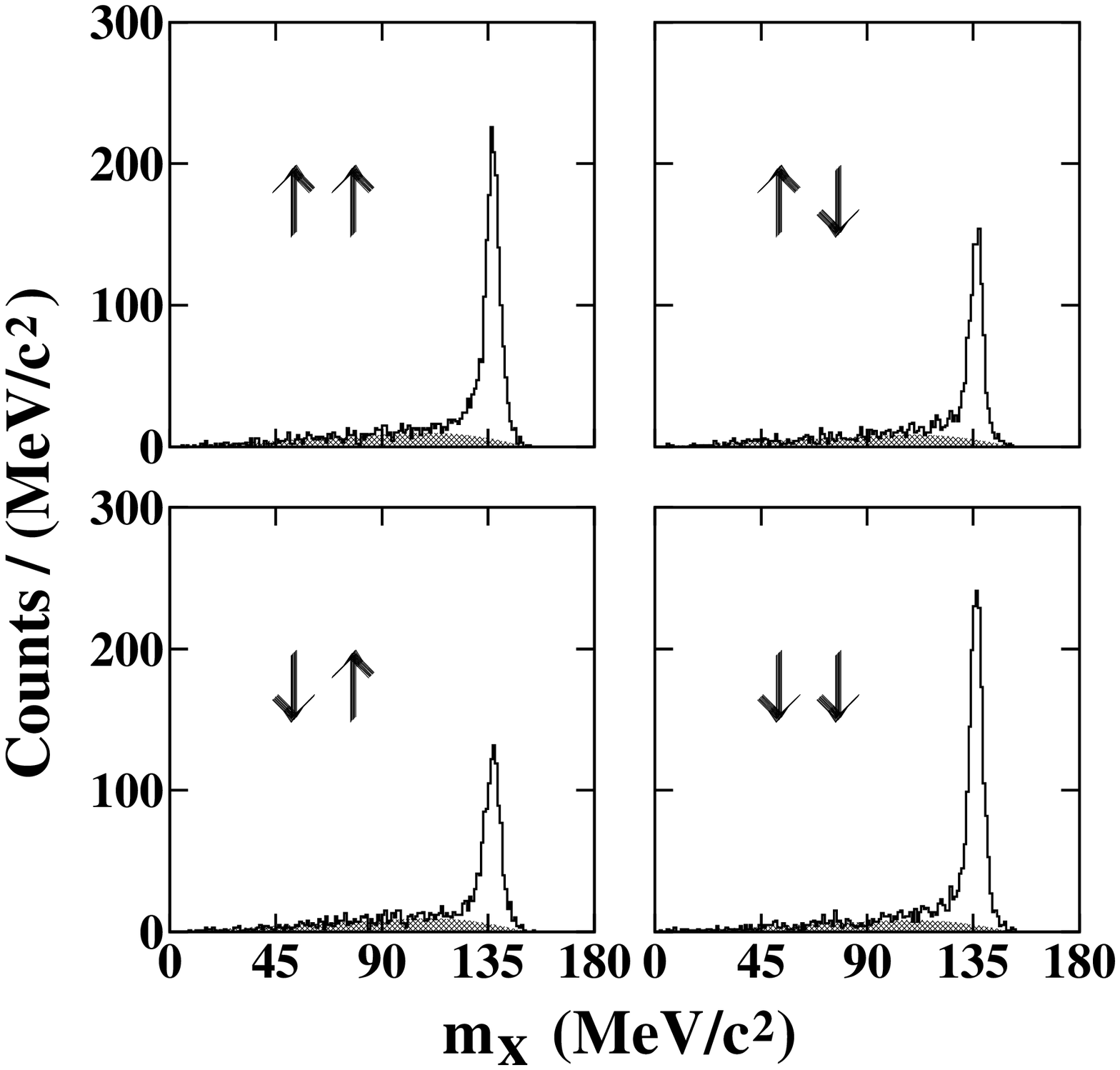}
}
\caption{(left) Forward detector and storage target cell, letters as
explained in the text. (right) Spinsorted distributions of the calculated 
missing mass ($m_x$) at $T_{beam}=325$ MeV. The arrows indicate
parallell or antiparallell beam and target polarization. 
The shaded area is the assumed background. \label{fig:TURKEY}}
\end{figure}
Particles emanating from near threshold reactions are peaked in the
forward direction in the laboratory system and thus a large acceptance
can be achieved with a relatively modest sized detector.
For measuring the four-momentum of the two outgoing protons
from $\overrightarrow{p} \overrightarrow{p}\rightarrow pp \pi^0$
four scintillator planes and two wire
chambers have been mounted within one 
meter downstream of the storage target cell (S), cf. Fig.~\ref{fig:TURKEY}. 
The maximum
$\theta _{lab}$ of the protons ranges from 22 to 33 degrees 
for bombarding 
energies in the interval 325 to 400 MeV. This angular range is fully
covered except at 400 MeV where there is a cut-off at the utmost
angles. The detectors are mounted symmetrically surrounding the beampipe 
which set the 
limit of the minimum detection angle to $\theta _{lab} \simeq 5 ^{\circ}$.

The four scintillator planes, with thicknesses of 
1.5 mm (F), 103 mm (E), 153 mm (K) and 6.4 mm (V), are 
all made of Bicron (BC408) 
and segmented into four (F, K) or eight sectors (E, V).
The outer radii are 203 mm (F), 369 mm (E), 423 mm (K) and 610 mm (V).
The position dependence of the light collection efficiency is
obtained from the data and is used to determine the absolute 
proton energies on an event-by-event basis.

For monitoring of luminosity and the beam and target polarization pp elastic  
scattering is detected in four additional scintillators (X) placed at 
$\theta _{lab} = 45 ^{\circ}$
at four azimuthal angles 
$\phi =\pm 45^{\circ}$ and $\phi =\pm 135 ^{\circ}$.
The thickness of these detectors is 51 mm. The product of beam 
and target polarization
($\rm P_B \cdot P_T$) is obtained by measuring the spin correlation
coefficient combination $A_{xx} - A_{yy}$ which is large
at $\theta _{lab} = 45$ degrees~\cite{PRZEWOSKI}.
The integrated luminosity $\int L dt$ is given by the known pp
elastic cross section.

Track reconstruction is based on information from the two wire chambers,
each with two orthogonal planes. The wire spacings are 3.175 mm and 6.35 mm, 
respectively. To avoid ambiguities where 2 prong events
hit the same wire the chambers are rotated $45 ^{\circ}$ with respect to each
other. The speciality of the design is that at the center of the 
sensitive area there is an opening 
that allows for the beam pipe to come through~\cite{SOLBERG}.

The trigger for an event
to be processed in the data acquisition as a 2 prong event is that one
or more F-segments and at least two 
E-segments  fire but the V-plane does not. For diagnostics and detector
tests single prongs are also recorded. 
The overlap between neighboring
scintillator segments, designed to avoid dead areas, is taken into account
in the matching between the particle's reconstructed track and which 
scintillator sectors that fired.

\section{Data analysis and Results}\label{sec:analysis}
For the primary event selection the following cuts and conditions are 
imposed to the data:
Particle identification is done by means of time-of-flight(E-F)
vs. the energy deposited in the E-detector (for protons that are
stopped in the E-detector) or by $\Delta$E/E plots, i.e. E vs. K 
(for protons passing through the E-detector). 
A cut is done in the reconstructed vertex spectrum of the event, which
shows the extended target of the storage cell.
An angular 2D cut is made for the opening angle between the two protons
vs. their difference in $\phi$.
The four wire chamber signatures that are processed are the
{2222}, {2221}, {2223}, {2233} multiplicities, where each number denotes the 
number of clusters in the four planes (ideally always two). 
The parameters of two straight lines passing through a common origin
on the beam axis is fitted in an iterative process to the input wire
positions, and a cut is made in the spectrum of the $\chi ^2$ of the 
adopted best fit.

\subsection{Energy calibration}
The light (corrected for position dependence) collected in the 
stopping scintillators E and K are added together. The resolution of the 
$\pi ^0$ missing mass peak determines the multiplicative constant
of an assumed linear calibration (the effect of a second degree 
correction is negligible for the observables studied), and the 
offset is decided such that the unpolarized $\theta$ distribution 
of the $\pi ^0$ in the c.m. system is symmetric. The constants 
are correlated and thus determined simultaneously. 

\subsection{Systematic errors}

The effects on the observables by varying the event selection criteria
were studied in a systematic way. The results obtained for 
reasonable variations were all within statistical errors and were averaged
over in the final result. 

One non-negligible effect was the error 
originating from the amount of background subtraction. 
It was estimated by varying
the integration limits of the $\pi ^0$ peak, and added in quadrature
to the statistical error. Particles that go undetected into the beam 
pipe represent a potential systematic error. The final-state
interaction between the two protons is only prominent in a 
relative S-wave and thus a bias is introduced in the ratio
to the other partial waves. In the analysis the $\Delta \sigma _T$ was
studied as a function of an artificial $\theta_{lab}$ cut-off using 
an effective-range expansion for the pp final-state
interaction~\cite{MEY92}. A fit was made and extrapolated to $0 ^{\circ}$. 
The correction was the size of an error bar. The effect from the hole
on the other observables was found to be negligible.

\subsection{Results}

The missing-mass spectrum of the undetected $\pi ^0$
was sorted according to beam and target spin combinations
in order to determine $\Delta \sigma _T$, defined
as the difference of the two cross sections
 $\sigma (\uparrow \downarrow + \downarrow \uparrow ~) - 
\sigma (\uparrow \uparrow - \downarrow \downarrow)$
, where the arrows 
denote $\pm y$ polarization of beam and target, respectively, 
cf. Fig.~\ref{fig:TURKEY}.

The spin dependent cross section (see Appendix A of Ref.~\cite{MEYDIAG}) can,
for the non-zero beam and target
polarization components in the present experiment, 
be written as
\bea
\sigma (\overline{P}^B, \overline{P}^T)=\sigma_0[1+
(P_y^B A_y^B +P_y^T A_y^T)\cos\phi-P_x^T A_y^T \sin\phi \nonumber \\
+ P_y^B P_x^T\frac{A_{xx}-A_{yy}}{2}\sin 2\phi 
+ P_y^B P_y^T (\frac{A_{xx}+A_{yy}}{2} - \frac{A_{xx}-A_{yy}}{2}\cos 2\phi)].
\eea
where $\sigma_0$ is the unpolarized cross section, 
$\phi$ is the azimuthal angle of the $\pi^0$, $P_{x(y)}^{B(T)}$ denote the
horizontal (vertical) polarization of beam (target) and 
$A_y^{B(T)}$ is the angle-integrated analyzing power related to the 
beam (target).
For $\overrightarrow{p} \overrightarrow{p}\rightarrow pp \pi^0$
we have $A_y^B \equiv - A_y^T$. 

\begin{figure}[htb]
\epsfxsize=10cm
\epsfysize=4.2cm
\epsfbox{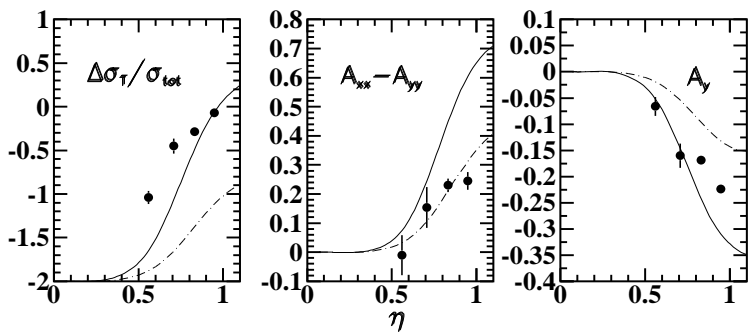}
\caption{The observables $\Delta \sigma _T /\sigma_{tot}$,
  $A_{xx}-A_{yy}$ and $A_y$ as a function of $\eta$, the maximum
  momentum of the pion in the c.m. system divided by 
$m_{\pi}$. The full-drawn line represents the full model as described 
in the text, the dashed-dotted line is calculated excluding the 
$\Delta$.\label{fig:DATA} }
\end{figure}

The expected $\phi$ dependencies were fitted to the 
spin-sorted experimental $\phi _{\pi ^0}$ distributions 
corresponding to the analyzing power $A_y$ and the spin
correlation coefficient combinations $A_{xx} - A_{yy}$
and $A_{xx} + A_{yy}$, (the latter identical to 
$-\Delta \sigma _T/\sigma_{tot}$). The results for the spin observables 
are given in Ref.~\cite{MEYER98} and shown in Fig.~\ref{fig:DATA}.
Also depicted are curves according predictions by a 
microscopic model calculation made by 
C. Hanhart et al.,
which is based on direct production and off-shell rescattering.
Contributions from HME ($\omega$) are added
and the $\Delta$ isobar is included 
explicitly~\cite{HANHART98}$^,$~\cite{HANHARTPRIV}.
\section{Conclusion}
We have described an experimental set-up and the first useful
measurements of spin observables of pion 
production below a bombarding energy of
400 MeV. Theoretical predictions for these observables are still scarce
and at variance with our data~\cite{HANHART98}.
Recent new activity in the field will presumably lead to further 
understanding of the production mechanisms for near threshold pion production.
In progress with the same set-up are also measurements 
with a polarized beam with both y- and z-components in combination
with a target polarized in x-, y- and z-directions. 
Thus we will be able to determine $\Delta \sigma _L$, $A_{xz}$ and
$A_{zz}$ and finally to deduce in a model-independent way the 
contributions of individual partial waves in this energy domain.

\end{document}